\documentclass[12pt]{article}
\usepackage{amsmath,amssymb,amscd}
\usepackage{epsfig}
\begin{document}

\begin{center}
{\bf MARKET MILL DEPENDENCE PATTERN IN THE STOCK MARKET: MULTISCALE CONDITIONAL DYNAMICS}
\end{center}


\begin{center}
\bf{ \large Sergey Zaitsev$^{(a)}$, Alexander Zaitsev$^{(a)}$, \\Andrei
Leonidov$^{(b,a,c)}$\footnote{Corresponding author. E-mail leonidov@lpi.ru}$^,$\footnote{Supported by the RFBR
grant 06-06-80357}, Vladimir Trainin$^{(a)}$, }
\end{center}
\medskip

(a) {\it Letra Group, LLC, Boston, Massachusetts, USA}

(b) {\it Theory Department, P.N.~Lebedev Physics Institute, Moscow, Russia}

(c) {\it Institute of Theoretical and Experimental Physics, Moscow, Russia}



\bigskip

\begin{center}
{\bf Abstract}
\end{center}

Market Mill is a complex dependence pattern leading to nonlinear correlations and predictability in intraday
dynamics of stock prices. The present paper puts together previous efforts to build a dynamical model reflecting
the market mill asymmetries. We show that certain properties of the conditional dynamics at a single time scale
such as a characteristic shape of an asymmetry generating component of the conditional probability distribution
result in the "elementary" market mill pattern. This asymmetry generating component matches the empirical
distribution obtained from the market data. We discuss these properties as a mixture of trend-preserving and
contrarian strategies used by market agents. Three basic types of asymmetry patterns characterizing individual
stocks are outlined. Multiple time scale considerations make the resulting "composite" mill similar to the
empirical market mill patterns. Multiscale model also reflects a multi-agent nature of the market.

\newpage

\section{Introduction}

The present paper continues a series of papers studying the complex dependence patterns in high frequency stock
price dynamics \cite{LTZ05,LTZZ06a,LTZZ06b,LTZZ06c,LTZZ07}. The most important of them, the market mill
asymmetries \cite{LTZZ06a,LTZZ06b,LTZZ06c,LTZZ07}, correspond to specific probabilistic interrelations between
consequent price increments. The term "market mill" refers to a mill-like asymmetric four-blade dependence pattern
\cite{LTZZ06a}, see Fig.~1.  The main emphasis of \cite{LTZZ06a,LTZZ06b,LTZZ06c} was on systematic
phenomenological description of the market mill asymmetries and other related properties of high frequency stock
price dynamics.

In \cite{LTZZ07} a causal conditional dynamics model leading to the market mill asymmetries and nonlinear
dependence of expectation value of a future price  increment y ("response") on the value of a realized price
increment x ("push") was suggested. The model described probabilistic relation between  the push $x$ and the
response $y$ in terms of the three - component conditional distribution ${\cal P}(y|x)$. The distribution ${\cal
P}(y|x)$ was described as an $x$-dependent additive superposition of the symmetric contribution ${\cal P}^0(y|x)$
and the asymmetry-generating components ${\cal P}^+(y|x)$ and ${\cal P}^-(y|x)$ characterized by a bias towards
trend-preserving and contrarian strategies correspondingly. The model of \cite{LTZZ07} referred to a single time
scale.

It is however well known that a description of certain features of stock price dynamics requires accounting for
multiple time scales, at the level of both price increments (returns) \cite{Man,G80,LB01,LZ03,LB05,LB06,LBY07,P08}
and microscopic long-memory properties of order flow and trades \cite{LF04,BGPW04,BKP06}. In particular, the
presence of several distinct time scales in volatility dynamics was explicitly demonstrated in \cite{LZ03}. In
\cite{LTZZ06a} the empirical market mill patterns were shown to exist at different time scales ranging from
minutes to hours.

In the present paper we incorporate the idea of multiple time scales into the market mill model. First we
introduce an elementary market mill mechanism at a fixed time scale. We describe an easier way of specifying the
elementary market mill by reformulating the model of \cite{LTZZ07}  in such a way that $y$ is a {\it sum} of noise
and non-random asymmetry generating components. Introducing specific features of the non-random component based on
empirical data we come up with the market mill pattern. Then we build a multiscale composite mill as a weighted
superposition of elementary asymmetry-generating mechanisms operating at different timescales.

The outline of the paper is as follows.

We start with a description of generic features of the market mill asymmetries in paragraph {\bf 2.1}. Particular
emphasis is put onto formulating a version stock price dynamics with an additive superposition of noise and
asymmetry-generating mechanisms.  In paragraph {\bf 2.2} we describe the conditional distribution allowing to
reproduce all observed market mill asymmetries. The properties of explicit dynamical model giving rise to a single
time scale elementary market mill asymmetries are discussed in paragraph {\bf 2.3}. The composite multiscale
dynamics allowing to reproduce all the properties of the market mill asymmetries is described in paragraph {\bf
2.4}. The section {\bf 3} contains a discussion of the origin of the market mill asymmetries in terms of three
basic strategies, market mill, trend-following and contrarian, used by market participants. We demonstrate that
appropriately weighted superpositions of these basic strategies allows to describe various two-dimensional
asymmetry patterns characterizing individual stocks.  We formulate our conclusions in section {\bf 4}.

\section{Conditional dynamics}

\subsection{Qualitative features}

At the fundamental level the ultimate goal of studying the dependence patterns in price dynamics is to describe
the observed dependence patterns between price increments $x=p(t_1+\Delta T)-p(t_1)$ and $y=p(t_2+\Delta
T)-p(t_2)$, where $t_2 \geq t_1+\Delta T$, in terms of explicit strategies used by market participants. These
strategies are realized through systematic reaction of market participants to the information about the sign and
magnitude of the increment $x$ leading to some predictability of the increment $y$. A strategy is thus fully
described by probabilistic properties of $y$ at given $x$ i.e. by a conditional distribution ${\cal P}(y|\,x)$. An
existence of dependence patterns relating the push $x$ and the response $y$ is then reflected in systematic
non-random features of ${\cal P}(y|\,x)$ which induce, in turn, dependence patterns of the full bivariate
distribution ${\cal P}(x,y)$ such as the market mill ones.

Let us describe the most important properties of ${\cal P}(y|\,x)$ following from an analysis of empirical data.
To this aim let us introduce an additive decomposition of the response $y$ into noise $y_{\rm rand}$ and
systematic $y_{\rm mill}$ contributions:
\begin{equation}\label{sumresp}
    y(x) \, = \, y_{\rm rand}(x) + y_{\rm mill} (x) \, ,
\end{equation}
where the random component $y_{\rm rand}(x)$ is described by a distribution ${\cal P}_0 (y_{\rm rand}|\,x)$ and
the systematic component $y_{\rm mill}$ is described by a distribution ${\cal P}_{\rm mill}(y_{\rm mill}|\,x)$.
The probability distribution for $y(x)$ is, naturally, ${\cal P}(y|\,x)$. A graphic illustration of the
decomposition (\ref{sumresp}) and qualitative features of the corresponding probability distributions ${\cal P}_0
(y_{\rm rand}|\,x)$, ${\cal P}_{\rm mill}(y_{\rm mill}|\,x)$ and ${\cal P}(y|\,x)$ is given, for the positive push
$x$, in Fig.~2 in columns {\rm I}, {\rm II} and {\rm III} respectively. Let us note that the above additive
decomposition is very natural from the point of view of an agent-based description where the orders originating
from different strategies are {\it added} in the time interval under consideration at each evolution step.

The distribution ${\cal P}_0 (y_{\rm rand}|\,x)$ is a symmetric function of its argument, ${\cal P}_0 (y_{\rm
rand} |\,x)={\cal P}_0(-y_{\rm rand}|\,x)$, see column {\rm I} in Fig.~2, and describes the main contribution to
the conditional dynamics. Market data shows \cite{LTZZ06b,LTZZ07} that the relative weight of the symmetric
contribution is dominant with respect to the asymmetric one\footnote{A quantitative analysis of the relative
weight of the asymmetric contribution can be found in \cite{LTZZ07}.}.

On top of the dominating random dynamics described by ${\cal P}_0 (y_{\rm rand}|\,x)$ there exists a systematic
asymmetry-generating dynamics described by ${\cal P}_{\rm mill}(y_{\rm mill}|\,x)$ illustrated in column {\rm II}
of Fig.~2 for $x=x_0>0$. Its message can be formulated in a simple statement. If the push $x$ is followed by the
response $y$ that does not exceed $x$ by magnitude, it is slightly more probable that the response is contrarian,
${\rm sign}(y)=-{\rm sign}(x)$. For responses with the magnitude exceeding that of the push it is, on contrary,
more probable that the response is trend-following, ${\rm sign}(y)={\rm sign}(x)$. For example, for positive
pushes $x>0$ there is a small bias towards responses of two sorts: those in the interval $y \in [-x,0]$ and those
in the interval $y>x$. Let us note that this particular property of ${\cal P}_{\rm mill}(y_{\rm mill}|\,x)$ also
goes along with the empirically observed features, see Fig. 3.

The above-described coexistence of dominating symmetric and subdominant asymmetric responses leads to the total
distribution with the shape sketched in column {\rm III} in Fig.~2. Technically the conditional distribution
${\cal P}(y|\,x)$ is given by an appropriate convolution of the distributions of the random and systematic
components
\begin{equation}\label{condistot}
 {\cal P} (y|\,x) = \int dy_{\rm rand} dy_{\rm mill} \, \delta (y - y_{\rm rand}-y_{\rm mill}) \,
 {\cal P}_0 (y_{\rm rand}|\,x) \, {\cal P}_{\rm mill} (y_{\rm mill}|\,x)
\end{equation}

With the specified probabilistic model for the conditional distribution ${\cal P}(y|\,x)$ the last step is to
establish its relation to the full bivariate distribution ${\cal P}(x,y)$ and its asymmetry patterns. Let us
remind that the specific dependence patterns of major interest to us, the market mill asymmetries
\cite{LTZZ06a,LTZZ06b,LTZZ06c}, refer to specific asymmetries of the bivariate distribution ${\cal P}(x,y)$ with
respect to the axes $x=0$, $y=x$, $y=0$ and $y=-x$.

First, one reconstructs the bivariate distribution ${\cal P}(x,y)$ from conditional dynamics described by ${\cal
P}(y|\,x)$. This reconstruction is based  on the relation ${\cal P}(x,y)={\cal P}(y|\,x) {\cal P}(x)$, where
${\cal P}(x)$ is a corresponding marginal distribution.

The market mill asymmetry patterns are described in terms of corresponding nontrivial asymmetric components ${\cal
P}^a (x,y)$ of the distribution ${\cal P}(x,y)$. For example, for the asymmetry with respect to the axis $x=0$
this asymmetric component reads
\begin{equation}
{\cal P}^a \, = \, \frac{1}{2} \left( {\cal P} (x,y) - {\cal P}(x,-y)  \right )
\end{equation}
The corresponding market mill dependence pattern refers \cite{LTZZ06a} to the specific shape of ${\cal P}_{\rm
mill} (x,y) \equiv {\cal P}^a (x,y) \cdot \Theta \left[ {\cal P}^a (x,y) \right]$, where $\Theta$ is the Heaviside
step function. The asymmetric components  ${\cal P}^a (x,y)$ are also responsible for nontrivial conditional
dependencies between $y$ and $x$. For example, again in the case of the asymmetry with respect to the axis $x=0$,
the mean conditional response $\langle y \rangle_x$\,,
\begin{equation}\label{condmeanpush}
\langle y \rangle_x \, = \,  \int dy \, y \,  \frac{{\cal P}^a (x,y)}{{\cal P}(x) } \, ,
\end{equation}
has a specific $z$-shaped dependence on $x$ \cite{LTZZ06a}.

The required identification of noise and asymmetry-generating components of ${\cal P}(y|\,x)$ is achieved
\cite{LTZZ06a} by extracting, in complete analogy with the above-described procedure for the bivariate
distribution ${\cal P} (x,y)$, the symmetric and mill components of ${\cal P}(y|\,x)$. One could say that  ${\cal
P}_{\rm mill} (y_{\rm mill}|\,x)$ contains "irreducible" information on the asymmetry-generating dynamics. The
experimentally measured mill component \cite{LTZZ06a} is shown in Fig.~3. The explicit version of conditional
dynamics described in the next paragraph is based on stylized features of ${\cal P}_{\rm mill}(y_{\rm mill}|\,x)$
following from the analysis of Fig.~3 and illustrated in Fig.~4. Let us stress once again that from the shape of
the mill component shown in Figs.~3,4 it is clear that the systematic asymmetry-generating contribution $y_{\rm
mill} (x)$ originates from {\it both} trend-preserving and contrarian order placement.

\subsection{Quantitative formulation}

Let us now turn to a step-by-step description of the conditional dynamics in question.

A compact description of the conditional distributions shown in Figs.~2-4 can be given by dividing the $x-y$ plane
into eight sectors \cite{LTZZ06a} shown in Fig.~5  and introducing an indicator function $f_{\rm mill}(x,y)$ equal
to $1$ in the even sectors and equal to $0$ in the odd ones, see Fig.~5(a)\footnote{An explicit expression for
$f_{\rm mill}(x,y)$ is given in the Appendix.}. Then
\begin{equation}\label{conmilldist}
 {\cal P}_{\rm mill} (y_{\rm mill}|\,x) \, = \, f_{\rm mill}(x,y_{\rm mill}) \, {\cal P}_0 (y_{\rm mill}|\,x)
\end{equation}
where  ${\cal P}_0 (y|\,x)$ is a symmetric distribution of the response $y$ at some fixed push $x$. The indicator
function $f_{\rm mill}(x,y)$ cuts from   ${\cal P}_0 (y|\,x)$ the pieces having support on the corresponding
segments of the $y$ axis. These cuts and the corresponding support intervals in the $y$ axis are shown, for $x=\pm
\$ \, 0.07$, in Fig.~5(b). The lines in the shaded areas correspond to the segments of the $y$ axis carrying
nonzero contribution to ${\cal P} (y|\,x)$ in Fig.~3.

We see that the structure of systematic response $y_{\rm mill}$ depends, in an essential way, on the sign of the push $x$.

Let us first assume that the push $x$ is positive, $x>0$. The corresponding response $y_{\rm mill}$ could be
positive or negative. For $y_{\rm mill}>0$ its value lies in the interval $y_{\rm mill}>x$ with the probabilistic
weight determined by the part of distribution in Fig.~4 (a) marked with green. For $y_{\rm mill}<0$ its value lies
in the interval $y_{\rm mill} \in [-x,0)$ with the probabilistic weight determined by the part of distribution in
Fig.~4(a) marked with red.

In terms of systematic strategies of market agents this corresponds to a push-dependent mixture of
trend-preserving and contrarian strategies. Indeed, the part of distribution in Fig.~4(a) marked with green
corresponds to trend-preserving order placement favoring the price growth while that marked with red is contrarian
and corresponds to that favoring its decline.

If the push is negative, $x<0$, the response could again be negative or positive.  For $y_{\rm mill}<0$ its value
lies in the interval $y_{\rm mill}<x$ with the probabilistic weight determined by the part of distribution in
Fig.~4 (b) marked with green. For $y_{\rm mill}>0$ its value lies in the interval  $y_{\rm mill} \in (0,-x]$ with
the probabilistic weight determined by the part of distribution in Fig.~4(b) marked with red. Here again we see a
push-dependent mixture of trend-preserving (green) and contrarian (red) strategies.

The technical message of Figs.~2-5 is that a procedure of constructing \\${\cal P}_{\rm mill} (y_{\rm mill}|\,x)$
consists in cutting appropriate pieces from the symmetric distribution ${\cal P}_0(y|\,x)$.

In our simulations described below we shall assume, for simplicity,
that ${\cal P}_0 (y_{\rm mill}|\,x)$ does not depend on $x$\,\footnote{In reality the shape of  ${\cal P}_0
(y_{\rm mill}|\,x)$ does depend on $x$, see a detailed discussion in \cite{LTZZ06b}} and use a Laplace
distribution
\begin{equation}\label{dlaplace}
{\cal P}_0 (y) = \frac{1}{2 \sigma_{\Delta T}} \exp \left \{ -\frac{|\,y|}{\sigma_{\Delta T}} \right \}
\end{equation}
that gives a reasonably good description of the bulk of the distribution of price increments at small intraday timescales.

Then for small $|\,x| < \log 2 \cdot \sigma_{\rm \Delta T}$ a dominating strategy is the trend-preserving one
while for large $|\,x| > \log 2 \cdot \sigma_{\rm \Delta T}$ this is a contrarian one.

In what follows the asymmetry-generating distributions (\ref{conmilldist}) will be used for explicitly
constructing a series of price increments combining the noise $y_{\rm rand}$ and asymmetric $y_{\rm mill}$
contributions. In this construction a decision of using the systematic strategy is also randomized. In our
simulations we first generate the noise increment price series using the distribution (\ref{dlaplace}). Then we
move along this price series and, at each step, decide whether a nontrivial asymmetry-generating contribution will
be added to the noise increment in one of the following time intervals. Thus, in addition to knowing how to
generate the asymmetry-generating component as described by (\ref{conmilldist}),  we have to decide whether, for a
given realized price increment $x$, the asymmetric contribution is generated and define an interval in which the
systematic price increment will be added. It is convenient to first specify a target interval and then either
generate a nontrivial contribution to the increment in this interval or leave the interval's increment
untouched\footnote{The explicit formula describing the asymmetry-generating conditional dynamics is given in the
Appendix.}.

Randomization of an appearance of the asymmetry-generating component is realized by assuming that at each step the
asymmetry-generating contribution is either switched on with a push-dependent probability $\nu(x)$ or switched off
with a probability $1-\nu(x)$. If it is switched off a zero contribution $y_{\rm mill}=0$ is generated.

\subsection{Single-scale conditional dynamics. Elementary mill}

Let us now  describe an explicit numerical realization of the dynamical model described in the previous paragraph.
This model is an additive  superposition of a trivial memoryless dynamics generating uncorrelated price increments
sampled from the symmetric distribution ${\cal P}_0(y_{\rm rand})$ and the above-described nontrivial conditional
dynamics generating asymmetric response $y_{\rm mill}$ for a given push $x$ in the interval separated from that
corresponding to the push by a randomly chosen number of intervals $l$. The single-scale conditional dynamics
corresponding to the elementary mill is then fully specified by selecting a distribution ${\cal P}(l|L)$ in the
number $l-1$ of time intervals separating those corresponding to the push $x$ and the response $y_{\rm mill}$
where the parameter $L$ controls a shape of this distribution. In what follows we shall use ${\cal P}(l|L) \propto
\exp (-l/L)$ with $L=3$. More precisely, when reaching a time interval with the realized price increment $x$ in
it\footnote{Note that $x$ may contain asymmetric contributions generated at earlier steps.}:
\begin{enumerate}
\item{If $x \neq 0$ in a fraction of $\nu_0=0.12$ cases an {\it additional} price increment $y_{\rm mill}$ sampled
 from ${\cal P}(y_{\rm mill}|\,x)$ is added to the preexisting increment in the interval at some distance $l \, \Delta T$
 from the interval with the realized increment $x$  where $l$ is sampled from ${\cal P}(l|L)$.}
\item{ If $x=0$ no action is taken. }
\item{ The basic symmetric distribution ${\cal P}_0 (y)$ is a Laplace one (\ref{dlaplace}) and the width of the
 "elementary" distribution $\sigma=\$ \, 0.02$ corresponds to the observable standard deviation of price increments
 for $\Delta T_0 = 1 \,{\rm min.}$.   }
\end{enumerate}

The resulting four mills corresponding to the asymmetries with respect to the axes $x=0$, $x=y$, $y=0$ and $y=-x$
are shown in Fig.~6. We see that the the model gives a very good description of all the four market mill
asymmetries.

From the analysis of market data we know that the market mill patterns are observed at different intraday time
scales. Because of the response delay built in into the model one expects that the 1-minute elementary mill will
propagate some millness to asymmetries measured at larger time intervals. To analyze this issue, let us again turn
to the division of the $x-y$ plane into eight sectors  \cite{LTZZ06a}, see Fig.~5(a), and introduce as a
quantitative measure of "millness" the quantity $\rho_{\rm mill}$ which is a relative difference between the
density of even and odd sectors in the domain $ \{ x,y \} \in [-\delta p^*,\delta p^*]$:
\begin{equation}
\rho_{\rm mill} \, = \, \frac{(n_8-n_1)+(n_2-n_7)+(n_6-n_3)+(n_4-n_5)}{n_{\rm tot}}
\end{equation}
where $\{ n_i \}$ are numbers of points generated within an $i$-th sector of this square. In our analysis we used
$\delta p^* = \, \$ \, 0.3$.

The mean millness $ \langle \rho_{\rm mill} \rangle$ and its standard deviation $\sigma(\rho_{\rm mill})$ computed
for the market data for 2000 stocks traded at NYSE and NASDAQ stock exchanges in 2004-2005 for a set of time
intervals $\Delta T = 1,3,6 \, {\rm min}$ are shown in the first two rows of Table 1. In this computation the
total set of stocks was randomly divided into 20 groups containing 100 stocks each. In this way we obtained 20
values for $\rho_{\rm mill}$ within each group. Their mean and standard deviation are the numbers shown in Table
1.

In our theoretical simulations we created, for each case considered, 2000 time series of the same length as in the
above-described market data which we also divide into 20 subsets containing 100 time series each. The
corresponding mean values and standard deviations characterizing theoretical millness are computed from 20 values
characterizing these 20 subsets.

In the third and fourth row of Table 1 we show the mean millness and its standard deviation for the elementary
market mill model. We see that, at variance with experimental observations, the original millness generated by the
elementary mill at the scale of $\Delta T=1 \, {\rm min}$ gradually weakens with growing time interval $\Delta
T$\footnote{The last two rows in Table 1 contain the results obtained for the composite mill, see paragraph {\bf
2.4}.}.

\bigskip

\begin{center}
{\bf Table 1.}
\medskip

\begin{tabular}{|c|c|c|c|c|}
  \hline
  Source & Quantity & $\Delta T=1 {\rm min}$ & $\Delta T=3 {\rm min}$ & $\Delta T=6 {\rm min}$ \\ \hline
  Market data & $ \langle \rho_{\rm mill} \rangle$ & 1.52 & 2.32 & 2.32 \\ \hline
  Market data & $\sigma \left( \rho_{\rm mill} \right)$& 0.07 & 0.08 & 0.10 \\ \hline
  Elementary mill& $ \langle \rho_{\rm mill} \rangle $ & 1.85 & 0.94 & 0.39 \\ \hline
  Elementary mill & $\sigma \left( \rho_{\rm mill} \right)$ & 0.02 & 0.02 & 0.04 \\ \hline
  Composite mill & $\langle \rho_{\rm mill} \rangle$ & 0.87 & 1.71 & 1.71 \\ \hline
  Composite mill & $\sigma \left( \rho_{\rm mill} \right)$ & 0.01 & 0.04 & 0.04 \\ \hline
\end{tabular}

\bigskip
 Table~1.  Values of mean millness and its standard deviation (both in percent) for market data, elementary and composite mills.
\end{center}

\subsection{Multiscale conditional dynamics. Composite mill}

The analysis in the previous section has shown that the observable initial growth and consequent constancy of
millness $\rho_{mill} (\Delta T)$ can not be achieved by considering a single elementary mill operating at some
"starting" scale $\Delta T_0$. A natural generalization of this construction is achieved by augmenting conditional
dynamics corresponding to the elementary mill  by adding, with some weights, conditional dynamics mechanisms
operating at larger time scales. More explicitly, let us consider a sequence of time scales $\{ \Delta T_i=i \,
\Delta T_0 \}$. Then at some given moment $t$ there is a probability $\nu_1$ of switching on an asymmetric
conditional dynamics at the scale $\Delta T_0$, a probability $\nu_2$ of switching on an asymmetric conditional
dynamics at the scale $2 \cdot \Delta T_0$, etc.

Let us consider as an example a superposition of two mills operating at time scales $\Delta T_0$ and $2 \cdot
\Delta T_0$.

For the first dynamics the push at time $t$ is a price increment $\delta P ([t-\Delta T_0,t])$ and the response is
generated in the interval $[t+l_1 \Delta T_0,t+(l_1 +1) \Delta T_0]$ where $l_1$ is sampled from ${\cal P}(l|L)$
and the standard deviation of the response is $\sigma_{\Delta T_0}$.

For the second dynamics the push at time $t$ is a price increment $\delta P ([t-2*\Delta T_0,t])$ and the response
is generated in the interval $[t+l_2 \Delta T_0,t+(l_2 +1) \Delta T_0]$ where $l_2$ is sampled from ${\cal
P}(l|2*L)$ and the standard deviation of the response is $\sqrt{2} \sigma_{\Delta T_0}$.

In the general case for the i-th mill component the push at time $t$ is a price increment $\delta P ([t-i*\Delta
T_0,t])$ and the response is generated in the interval $[t+l_i \Delta T_0,t+(l_i +1) \Delta T_0]$ where $l_i$ is
sampled from ${\cal P}(l|i*L)$ and the standard deviation of the response is $\sqrt{i} \sigma_{\Delta T_0}$.

The above-described probabilistic construction can be termed a composite mill, where composition refers accounting
for mill mechanisms operating on a set of increasing time scales $ \{ \Delta T_i \}$.

Let us consider a regularly decaying infinite series of weights $\nu_i$, $\nu_i=0.8 \cdot \nu_{i-1}$ with
$\nu_1=0.12$. This leads to well-defined market mill asymmetries akin to the ones shown in Fig.~6. The resulting
mean asymmetry measure $\langle \rho_{\rm mill} \rangle$ and its standard deviation $\sigma(\rho_{\rm mill})$
calculated on the same set of time scales as in the previous section are shown in the last two rows in Table 1. We
see that by including conditional dynamics mechanism operating at a set of timescales allows to reproduce the
observed dependence of the millness $\rho_{\rm mill}$ on the time scale $\Delta_t$. This constitutes a clear
evidence of the existence of multiscale conditional dynamics.

\section{Discussion}

In this section we are going to make interpretation of the proposed model by adding a specific market sense to the
model and discussing major possible reasons for the price evolution to result into such a complex pattern as the
market mill.

The idea behind the developed conditional dynamics model is that properties of price increment in some time
interval are probabilistically related to the behavior of price increments in one or several preceding time
intervals. The properties of a price increment at some given timescale are determined by a weighted superposition
of signals originating from events occurring at different timescales and separated from the interval under
consideration by time intervals of varying length. This mechanism underlies the multiscale properties of price
dynamics leading to the composite mill dependence patterns and leads to successful description of observable
strength of market mill asymmetry on different time horizons. Let us stress that our description is based on the
superposition of noise and signal where noise is dominant and signal obeys conditional dynamics.

The phenomenon of market mill asymmetry is most naturally understood in terms of an existence of multiple market
agents/strategies leading to probabilistic dependence of past on future. The two basic types of such strategies
are trend-following, resulting in positive correlation between past and future increments, and contrarian, leading
to negative correlation between them. From the properties of conditional dynamics described in Section 2, see also
\cite{LTZZ07}, it is clear that the market mill asymmetry patterns arise as a result of a specific finely tuned
balance between trend-preserving and contrarian tendencies. At this point it is important to recall that a
detailed analysis of the asymmetry patterns characterizing individual stocks \cite{LTZZ06c} shows that the
clear-cut market mill asymmetry is characteristic only for a certain subgroup of stocks and two other subgroups,
predominantly trend-preserving and predominantly contrarian, can be identified. This diversity of stable
individual asymmetry patterns can most naturally be described by an existence of three basic asymmetry generating
modes, the market mill, contrarian and trend-following. The corresponding modes are illustrated, for positive
push, in Fig.~7. An individual asymmetry pattern reflects a particular superposition of these signals reflecting
the proportions in which corresponding strategies are present in the trading activity for the considered stock.
Such generalized signal space indeed allows to reproduce the properties of individual asymmetry patterns. In
Fig.~8 we compare the empirical \cite{LTZZ06c}, Figs. 8.1a, 8.2a and 8.3a, and calculated, Figs. 8.1b, 8.2b and
8.3b, asymmetry patterns for the stocks DIS, HDI and DE having clear-cut correlated, market mill and
anticorrelated asymmetry patterns respectively. Let us stress that the model description involves, with specific
weights, all the three above-described fundamental signals. In particular, taking into account an admixture of
market mill pattern is crucial for correctly reproducing the form of equiprobability lines for correlated and
anticorrelated asymmetry patterns in Figs. 8.1a and 8.3a.

To establish a link to the agent-based picture described in \cite{LTZZ07}, let us make a standard assumption that
price increments  $y$  are proportional to the cumulative signed volume of orders $\Omega_y$ in the time interval
under consideration \cite{KK99,PGGS01,GGPS03}. The signal probability distribution can then be rephrased in terms
of conditional probabilities of placing orders with cumulative signed volume $\Omega_y$ based on an information on
the sign and magnitude of the set of realized price increment $\{x\}$. The resulting signal distribution of signed
price orders  ${\cal P} \left( \Omega_y | \, \{x\} \right)$ is thus determined by appropriately weighted
contributions described by ${\cal P}^{\rm mill} \left( \Omega_y | \, \{x\} \right)$, ${\cal P}^{\rm cor} \left(
\Omega_y | \, \{x\} \right)$ and ${\cal P}^{\rm acor} \left( \Omega_y | \, \{x\} \right)$ referring to market
mill, trend-following and contrarian contributions respectively.

\section{Conclusions}

A reformulation of the three-component model of \cite{LTZZ07} providing an additive separation of noise and
asymmetry-generating contributions is described. Specific shape of the asymmetry generating component of the
conditional probability distribution at a single time scale leads to the elementary mill pattern. A multiscale
conditional dynamics taking into account, in addition to the initial elementary mill, appropriately weighted
conditional dynamics mechanisms combining trend-preserving and contrarian strategies operating on a set of
increasing timescales is proposed. This composite model based on multiscale dynamics is shown to reproduce the
market data on the market mill asymmetry for a set of timescales  as well as three basics types of asymmetry
patterns characterizing individual stocks.

\newpage

\begin{flushleft}
\large{ {\bf Acknowledgement}}
\end{flushleft}

V.T. is very grateful to Victor Yakhot for many useful discussions on topics related to the present paper.

\begin{flushleft}
\large{\bf{Appendix}}
\end{flushleft}

The explicit expression for the appropriately normalized indicator function $f_{\rm mill}(x,y)$ reads
\begin{eqnarray}
f_{\rm mill} (x>0,y) & = & 2\cdot \left [ \, \theta (y_{\rm mill}) \, \theta (y_{\rm mill}-x) +
 \theta (-y_{\rm mill}) \, \theta(y_{\rm mill}+x) \, \right  ]  \\
f_{\rm mill} (x<0,y) & = & 2\cdot\left [ \, \theta (y_{\rm mill}) \, \theta (-y_{\rm mill}-x) +
 \theta (-y_{\rm mill}) \, \theta(-y_{\rm mill}+x) \, \right ] \nonumber
\end{eqnarray}

The conditional distribution taking into account the randomization of the yield of systematic strategies used in
the simulations reads
\begin{equation}
 {\cal P}_{\rm asym}(y_{\rm mill}|\,x) \, = \, (1-\nu(x)) \, \delta (y_{\rm mill}) + \nu(x) \,{\cal P}_{\rm mill} (y_{\rm mill}|\,x)
\end{equation}
where $\delta (y_{\rm mill})$ is the Dirac delta-function.  We used the following simple parametrization of
$\nu(x)$:
\begin{eqnarray}
 \nu(x) & = & \nu_0 \, , \,\,\,\,   |\,x| > 0 \nonumber \\
 \nu(x) & = & 0 \, , \,\,\,\,\,\,\,  |\,x| = 0
\end{eqnarray}

\begin{figure}[ht]
 \begin{center}
 \epsfig{file=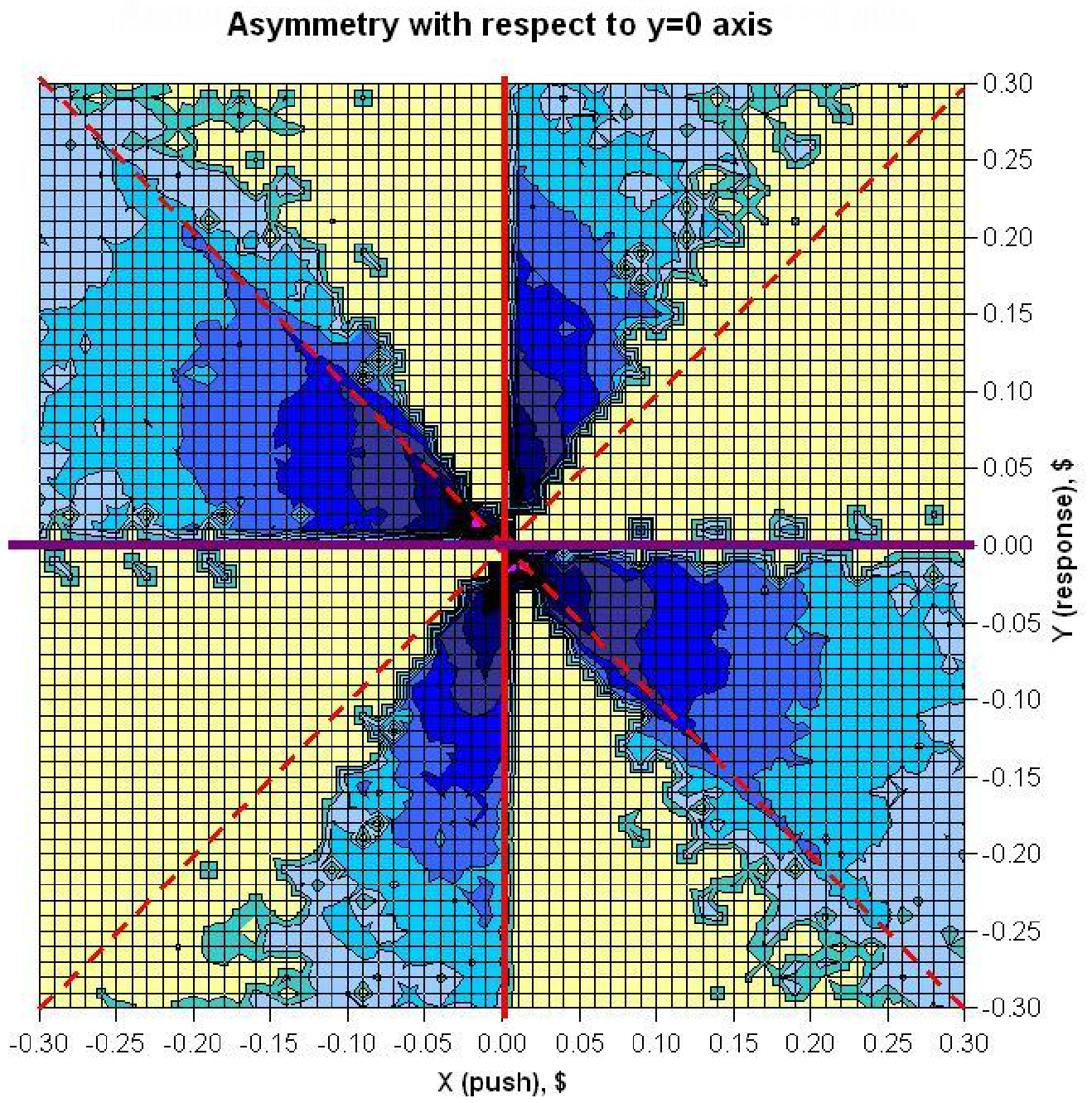,height=16cm}
 \caption{The market mill pattern, \cite{LTZZ07}.}
 \end{center}
\end{figure}

\begin{figure}[ht]
 \begin{center}
 \epsfig{file=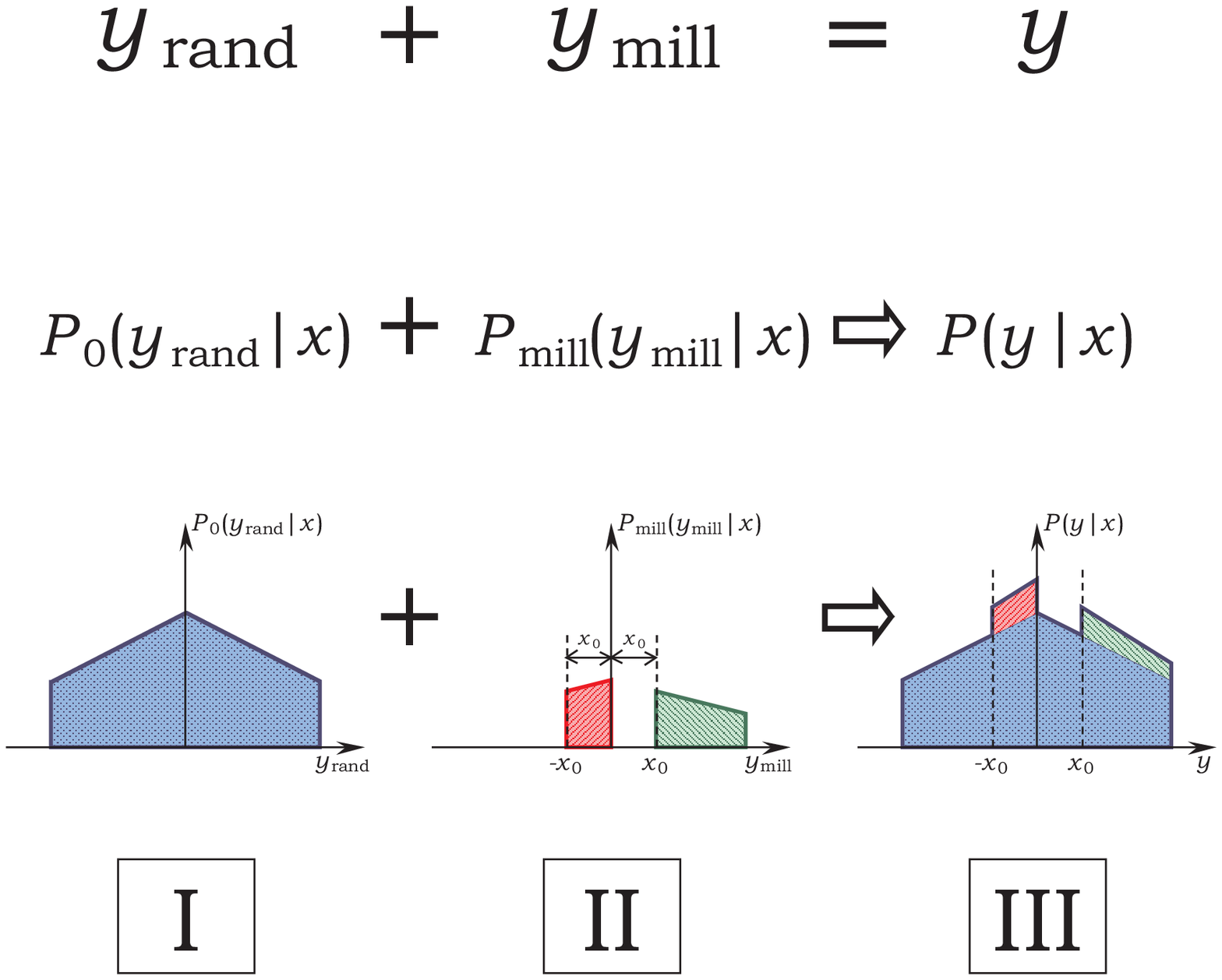,width=14cm}
 \caption{Sketch of the asymmetry-generating conditional dynamics}
 \end{center}
\end{figure}

\begin{figure}[ht]
 \begin{center}
 \epsfig{file=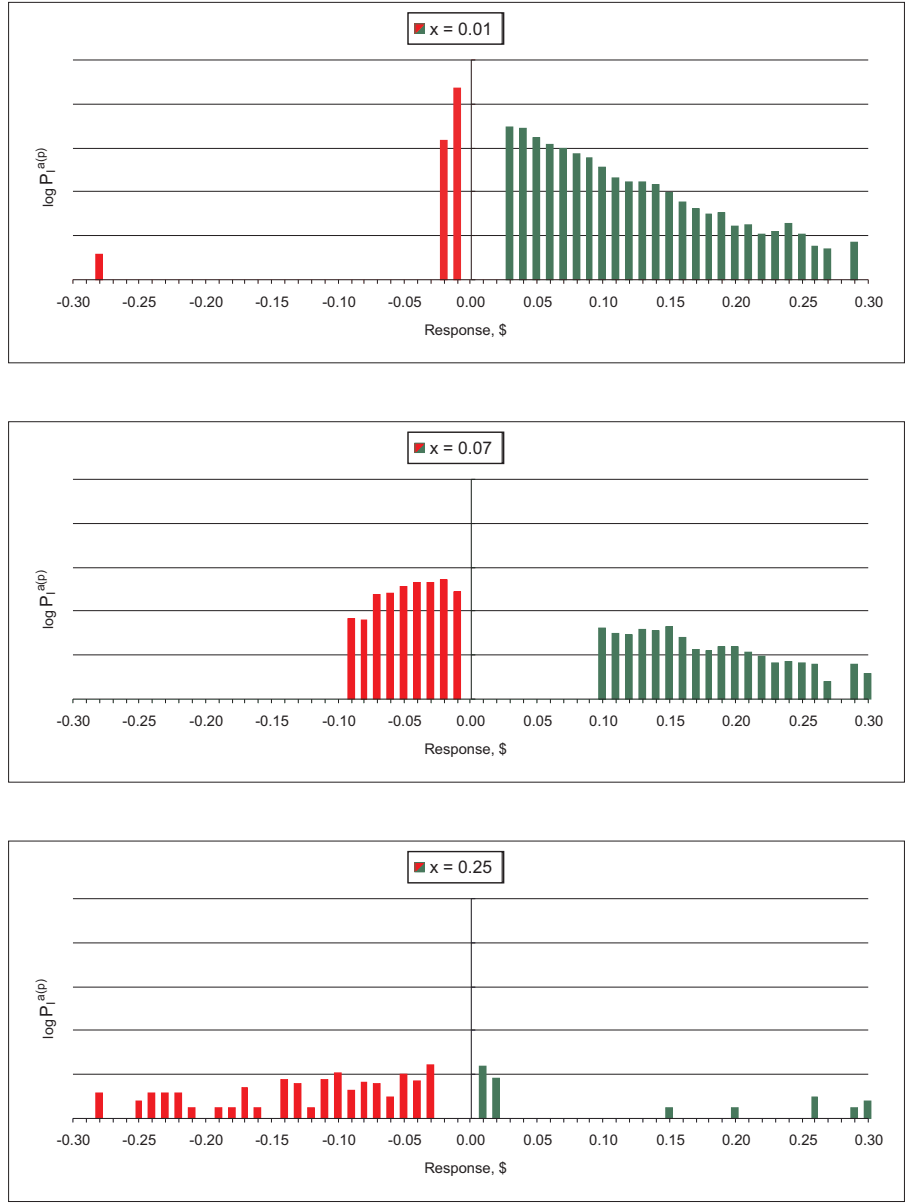,height=16cm}
 \caption{The observed mill distribution ${\cal P}_{\rm mill} (y_{\rm mill}|\,x)$ at different $x$ \cite{LTZZ06a}.}
 \end{center}
\end{figure}

\begin{figure}[ht]
 \begin{center}
 \epsfig{file=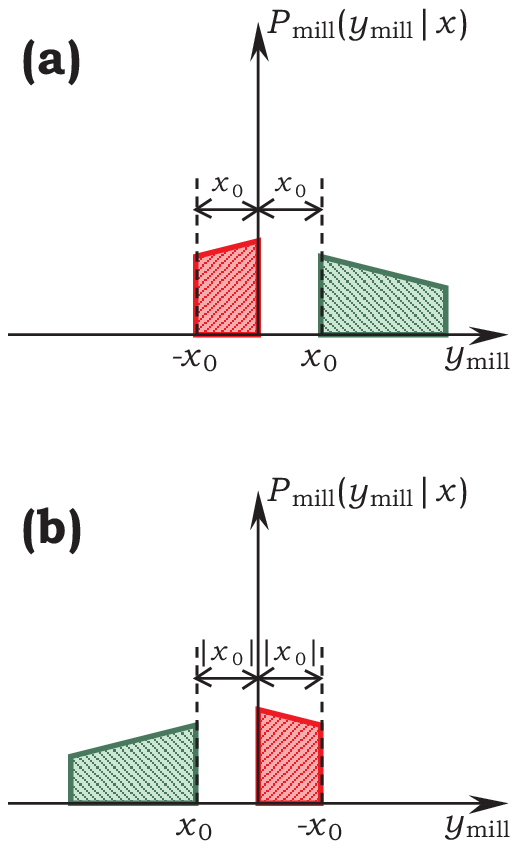,height=16cm}
 \caption{Asymmetry generating component of the conditional probability distribution
 ${\cal P}_{\rm mill} (y_{\rm mill}|\,x)$; (a) $x_0>0$; (b) $x_0<0$.}
 \end{center}
\end{figure}

\begin{figure}[ht]
 \begin{center}
 \epsfig{file=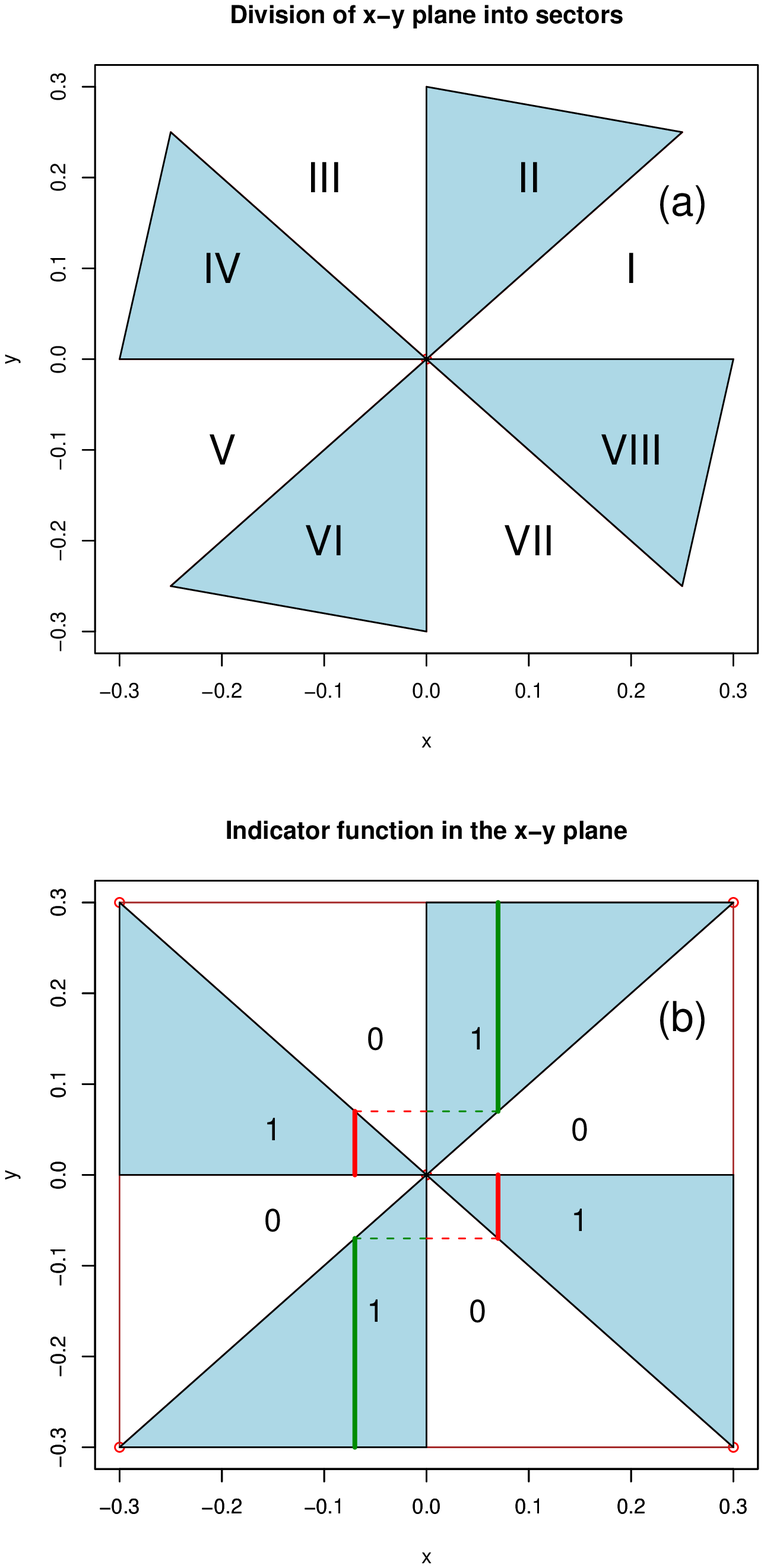,height=18cm}
 \caption{Sectors in $x-y$ plane \cite{LTZZ06a}.}
 \end{center}
\end{figure}

\begin{figure}[ht]
 \begin{center}
 \epsfig{file=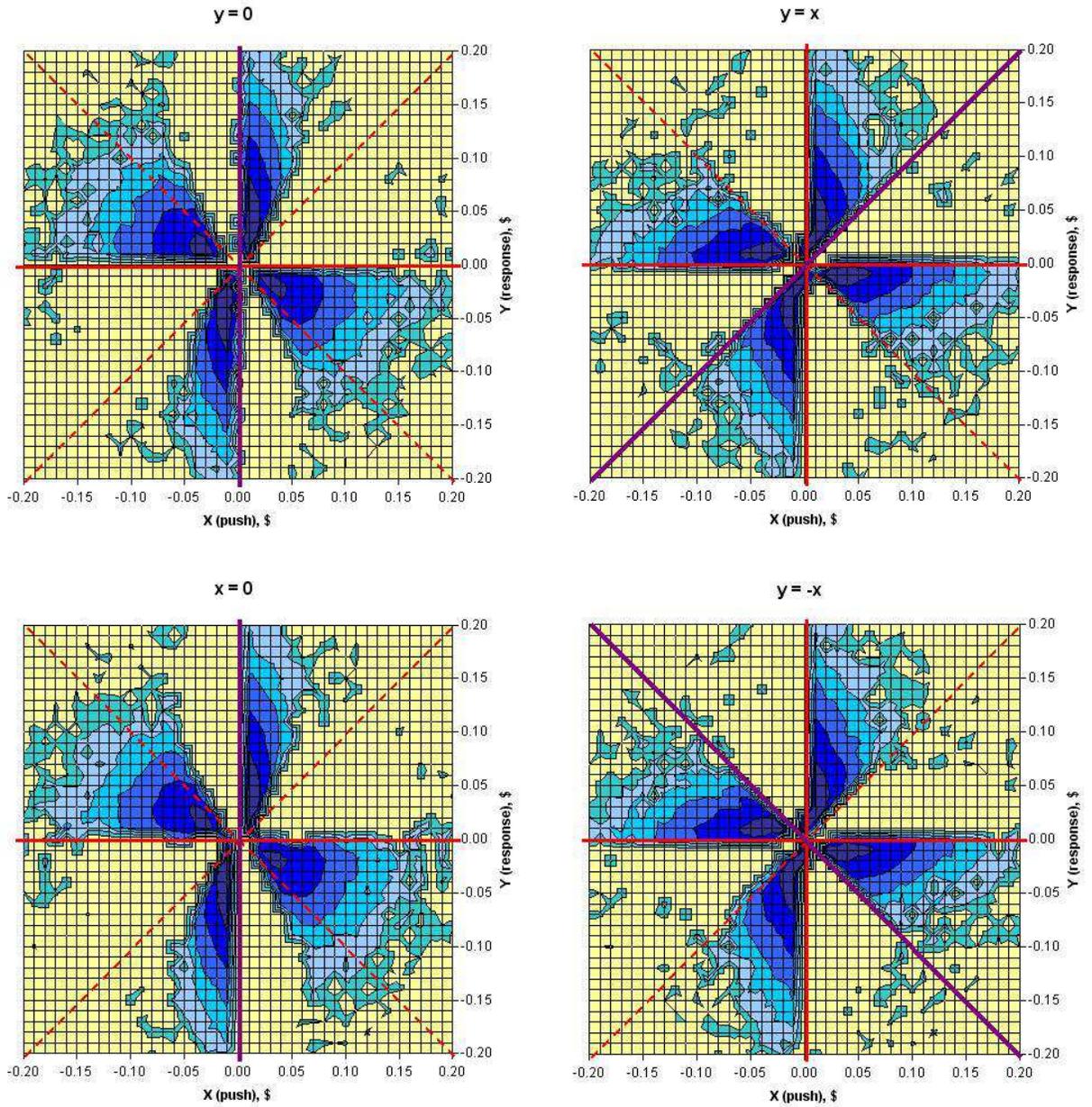,width=16cm}
 \caption{Model market mill patterns with respect to the axes $y=0$, $y=x$, $y=0$, $y=-x$, elementary mill, 
 $\Delta T=1 \, {\rm min}$.}
 \end{center}
\end{figure}

\begin{figure}[ht]
 \begin{center}
 \epsfig{file=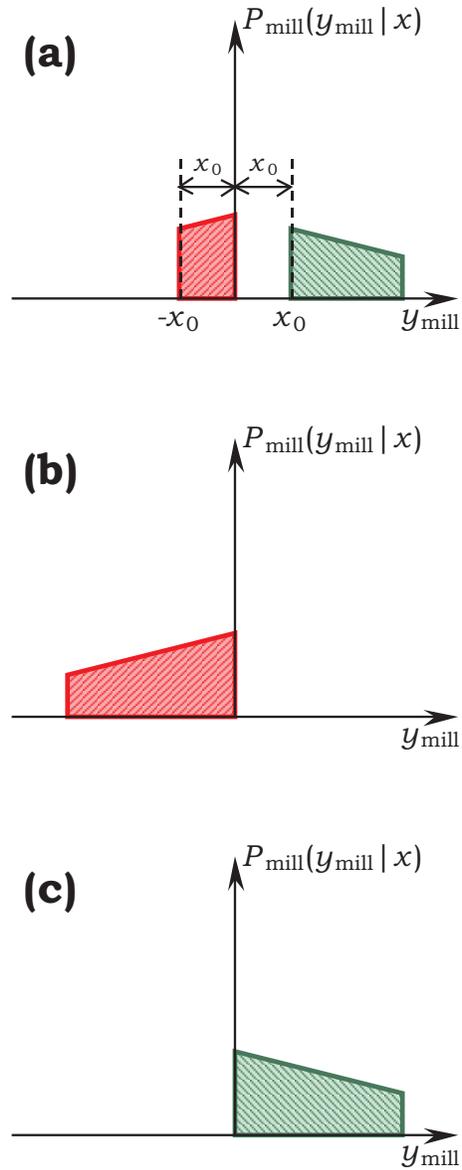,height=16cm}
 \caption{Three characteristic types of the asymmetry generating component of the conditional probability distribution:
  market mill (a), contrarian (b) and trend-following (c).}
  \end{center}
\end{figure}

\begin{figure}[ht]
 \begin{center}
 \epsfig{file=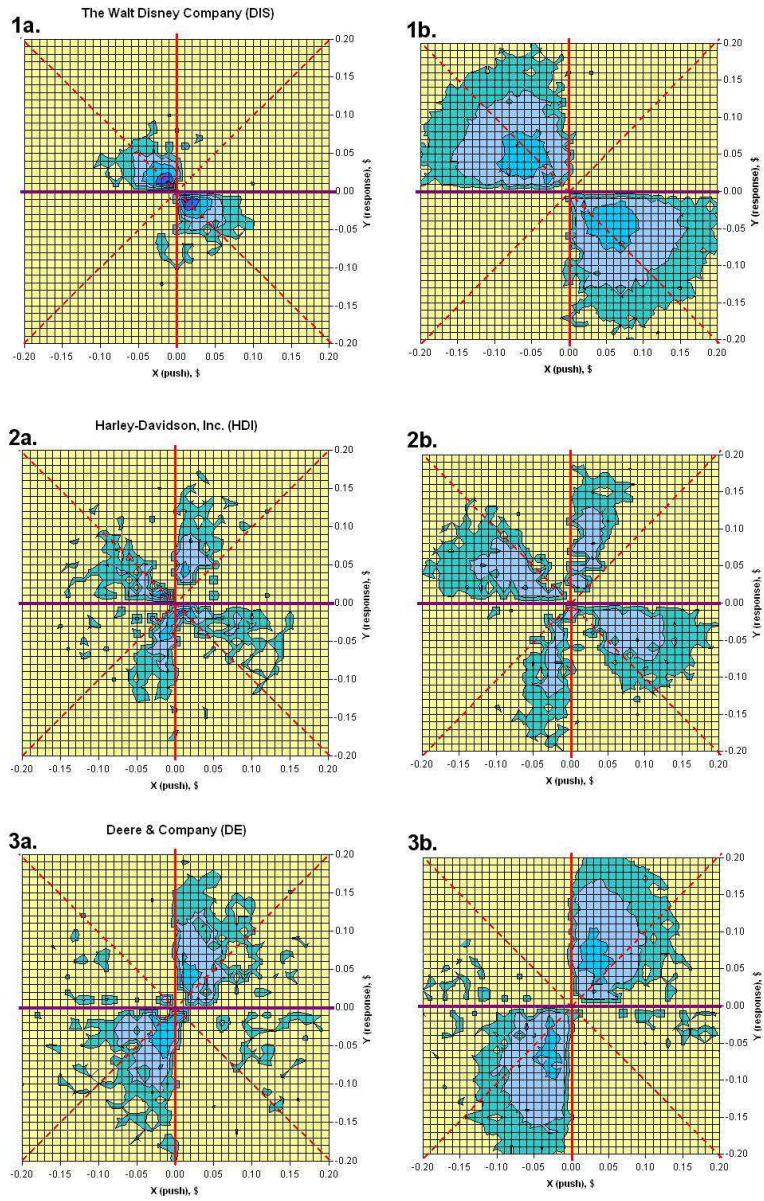,height=16cm}
 \caption{Model individual portraits (1b, 2b, 3b) versus market data (1a, 2a, 3a).}
 \end{center}
\end{figure}

\end{document}